\documentstyle[preprint,pra,aps,epsfig]{revtex}
\begin{document}
\draft
\title{\bf
Structure of Low-Energy Collective $0^{-}$-States in Doubly
Magic Nuclei
and 
Matrix Elements of the P-odd and P- and T-odd Weak Interaction
}
\author{O.K.Vorov, N.Auerbach, V.V.Flambaum
\footnote{Permanent address:
School of Physics, University of New South Wales, 
Sydney 2052, New South Wales, Australia}
}
\address{
School of Physics and Astronomy,
Tel Aviv University,
Ramat Aviv, 69978 Tel Aviv, Israel\\
and The European Center for Theoretical Physics, Trento, Italy
}
\date{19 March 1996}
\maketitle
\begin{abstract}
The structure of the collective 
low-energy $J^{\pi}=0^{-}$ (T=0 and T=1) modes 
is studied for a doubly magic nucleus in a schematic analytic model of 
RPA.
The $0^{-}$ phonon states 
($T= 0,1$) lie at energies 
$E_{T=0}(0^{-}) 
\alt \omega$ and 
$E_{T=1}(0^{-}) 
>
\omega$, where $\omega$ is the oscillator frequency.
The matrix elements of P-odd and P- and T-odd 
weak one-body potentials connecting the
ground state to these $0^{-}$-states, $W_{coll}$, are enhanced by the
factor 
$\sim 2 (\frac{\omega}{E
}
)^{1/2}A^{1/3} \sim 10$
as compared to the single-particle value $w_{sp}$ what can result in
values $|W_{coll}| \sim 20-30 eV$ 
if standard values of DDH parameters are used for $w_{sp}$.
Similar enhancement arises in the P- and T-odd 
case. 
\end{abstract}
\pacs{PACS: }

The purpose of this work is to study low-lying collective nuclear 
states
with quantum numbers $J^{\pi}=0^-$ and the matrix elements of 
parity nonconserving potentials connecting these states 
to the ground state.
The work is 
motivated not only by the interest in
the low-energy $0^-$ modes, which already attracted attention 
earlier in relation to pion (pre-)condensation, 
but also by the possibility of the collective 
enhancement of P- and P-,T-odd matrix elements 
of the pseudoscalar ``weak potentials'' in nuclei.
The importance of collective $0^-$ states for these effects
has been already revealed within doorway state approach
to the P-violation in compound nuclei \cite{A},\cite{AB}.
It was pointed out \cite{A} that the $0^-$ component of 
the spin-dipole state is an important mediator of parity 
mixing between nuclear states. 
This applies to both parity violating (P-odd) and time reversal 
conserving (T-even) as well as to P-odd and time reversal violating 
(T-odd) interactions.
The one-body potentials for these two symmetry violating
interactions can be approximately described by the operators
$(\vec{\sigma} \vec{p})$ 
for P-odd T-even and $(\vec{\sigma} \vec{r})$ for P-odd, T-odd.
When acting on a nuclear state both operators will excite
the $0^-$ mode. Since the weak interaction does not conserve 
isospin 
one considers both isoscalar ($T=0$) and ($T=1$)
$J=0^-$ spin-dipole states.

In Refs.\cite{FV1,FV2}, renormalization of the weak P- and P-,T-odd
potentials due to the residual strong interaction were calculated
and the positions of low energy $0^-$ modes was discussed.
The properties of $0^-$
($T=1$) states were studied earlier in an elaborated Hartree-Fock+RPA 
approach in Ref.
\cite{AK}.
Spin-dipole modes were studied in \cite{TCZ},\cite{AZ} within an
approximate RPA approach. 
An intersting analysis of related effects is 
presented in Ref.\cite{BJ}.

Here we show, using an approximate analytical RPA calculation,
that a realistic residual nucleon-nucleon interaction 
which has been widely used to describe various nuclear 
properties does indeed lead to formation of such collective 
$0^-$ modes. We give approximate quantitative predictions
for their characteristics.
We confine ourselve here to the case of doubly magic nucleus.
The density of states is relatively low.
We neglect in our study the coupling of the one-particle one-hole
$0^-$ states to all other configurations.

We write the nuclear Hamiltonian in the form
\begin{equation}
\label{HAMILTONIAN}
H = H_0 + V,
\end{equation}
where $V$ is the residual strong interaction
and $H_0$ is the singe-particle shell-model Hamiltonian 
of nucleons with  
momenta $\vec{p}_a$ and mass $m$ moving in a spherically
symmetric average potential $U(r)$ :
\begin{equation}
\label{HARMONIC}
H_0 = \sum_a \left( \frac{\vec{p}_a^2}{2m}+
U(r_a)
\right), 
\qquad 
U(r) = -|U_0| + \frac{m \omega^2 \vec{r}_a^2}{2} 
\end{equation}
(we use the harmonic oscillator well
of the depth $|U_0|$
and neglect for simplisity the shell splitting).

We employ the Landau-Migdal parametrization \cite{Migdal}
of the residual strong 
interaction chosing $V$ to be the sum of velocity 
independent $V_0$ and velocity dependent $V_1$ components,
$V= V_0 + V_1$.
The velocity independent term in the residual interaction 
is given by
\begin{equation}
\label{V0}
V_0(1,2)=C\delta(\vec{r}_1-\vec{r}_2)[f_{0}+f'_{0}\vec{\tau}_1
\vec{\tau}_2+
g \vec{\sigma}_1 \vec{\sigma}_2+
g' \vec{\tau}_1 \vec{\tau}_2 \vec{\sigma}_1
\vec{\sigma}_2],
\end{equation}
where 
$C=300$ $MeV \cdot fm^{3}$ is the universal Migdal constant.
(We use the convention $\tau_z(p)=-1,\tau_z(n)=1$ \cite{BM}). 

The spin- and momentum-dependent component of the interaction $V_1$
will be used in two forms: the first one can be written 
\cite{KS} in the 
form of a Landau-Migdal spin- and velocity dependent interaction,
\begin{eqnarray}
\label{V1}
V_1(1,2)= \frac{1}{4} C p_{F}^{-2} (g_{1}+g'_{1} \vec{\tau}_{1}
\vec{\tau}_{2})
(\vec{\sigma}_{1} \vec{\sigma}_{2})
[ \vec{p}_{1} \vec{p}_{2} \delta(\vec{r}_{1}-\vec{r}_{2}) +
\vec{p}_{1} \delta(\vec{r}_{1}-\vec{r}_{2}) \vec{p}_{2} + 
\nonumber\\
\vec{p}_{2} \delta(\vec{r}_{1}-\vec{r}_{2}) \vec{p}_{1} +
\delta(\vec{r}_{1}-\vec{r}_{2})\vec{p}_{1} \vec{p}_{2} ],
\end{eqnarray}
$p_F$ is the Fermi momentum.
In Eqs.(\ref{V0},\ref{V1}), the parameters $f$ and $g$ are the 
strength constants of order of unity (their values will be 
given below). Generally speaking, they contain $r$-dependence
which is usually parametrized via the density dependence.
However, this dependence is considerable only for the constants
$f$ and $f'$ which describe spin-independent components of the
interaction. 
In what follows, only spin-dependent components of the interaction
are essential, and the corresponding strengths $g_i$, $g'_i$ are  
density independent.
This phenomenological interaction can be viewed as a model of more
realistic but also more complicated forces. It is therefore convenient 
to use it first because of simplicity of calculations.
In fact, the results obtained for this
interaction manifest the correct tendencies (see also \cite{FV1}).

We also make calculations for the $\pi$- and $\rho$- meson exchange
interaction in its explicit form
\begin{equation}
\label{PIRHO}
V_{\pi+\rho} = 
\frac{f^2_{\pi}}{m^2_{\pi}} (\vec{\tau}_1 \vec{\tau}_2) \biggl[
(\vec{\nabla} \cdot \vec{\sigma}_1) 
(\vec{\nabla} \cdot \vec{\sigma}_2),
\frac{e^{-m_{\pi} r}}{ r} \biggl] 
- 
\frac{f^2_{\rho}}{m^2_{\rho}} (\vec{\tau}_1 \vec{\tau}_2) \biggl[
(\vec{\nabla} \times \vec{\sigma}_1) 
(\vec{\nabla} \times \vec{\sigma}_2),
\frac{e^{-m_{\rho} r}}{ r} \biggl] 
\end{equation}
using it instead of $V_1$ in Eq.(\ref{V1}).
Here, $m_{\pi}$ ($m_{\rho}$) are the $\pi$-meson ($\rho$-meson) mass,
$f_{\pi}$ and $f_{\rho}$ are the coupling constants, and 
$\vec{\nabla}$ is the derivative with respect to the nucleon
separation $\vec{r}= \vec{r}_1 - \vec{r}_2$;
$\times$ denotes the external vector product.
The standard ``Lorentz-Lorentz'' correction (see, e.g., \cite{JKWB})  
is assumed \cite{FV1}.

In the general case, the RPA equations can be written in the
form of equations of motion for the phonon creation operators 
$\hat{A}^{\dagger}$
\begin{equation}
\label{RPAequation}
- E_n \hat{A}^{\dagger}_n = 
[ \hat{A}^{\dagger}_n, 
{\cal H}_{RPA}
]
= [ \hat{A}^{\dagger}_n, H_0 ] + 
\langle [ \hat{A}^{\dagger}_n, V ] \rangle, 
\end{equation}
where the first equality comes from the commutator of the
phonon creation operator $\hat{A}^{\dagger}_n$ with the 
RPA
Hamiltonian 
${\cal H}_{RPA} \equiv const + 
\sum_n E_n \hat{A}^{\dagger}_n \hat{A}_n$
which is derived from the initial Hamiltonian $H_0 + V$
when the RPA phonons are used.
The expectation value 
in (\ref{RPAequation})
is evaluated using the uncorrelated ground state $|0^+\rangle$.
Within the RPA, the phonon operators are
\begin{equation}
\label{RPAphonon}
\hat{A}^{\dagger}_{n} = 
\sum_{im} X^{n}_{mi} \hat{c}^{\dagger}_m \hat{c}_i
- \sum_{im} Y^{n}_{mi} \hat{c}^{\dagger}_i \hat{c}_m
\end{equation}
where the $\hat{c}^{\dagger}_a$ and $\hat{c}_b$ are the 
operators of the creation and annihilation of nucleons
in the single-particle states marked by $a$ and $b$, respectively.
The amplitudes $X$ and $Y$ are related to the 
``forward'' and ``backward'' going graphs, respectively.
The index $n$ marks the modes, the total number of which ${\cal N}$,
coincides with the total number of the particle-hole pairs
with quantum numbers $0^{-}$.

The $0^{-}$ particle-hole states have energies 
$1 \omega, 3\omega, ...$ \quad. 
We are interested 
in low-energy collective states which should be dominated
by the $1 \omega$ transitions. 
The corresponding constituents of
the phonon modes can be conveniently parametrized 
by the elementary correlated $0^{-}$ particle-hole operators
$\hat{A}^{\dagger}_{0}$ and $\hat{A}_{0}$ defined below.
We use the following representation for the 
{\it collective}
phonon creation and destruction 
operators $\hat{A}^{\dagger}_{coll}$ and $\hat{A}_{coll}$
\begin{equation}
\label{COLLECTIVEphonon}
\hat{A}^{\dagger}_{coll} =
x \sum_{mi} (\vec{\sigma} \vec{a}^{\dagger})_{mi} 
\hat{c}^{\dagger}_m \hat{c}_i
- y   \sum_{mi} (\vec{\sigma} \vec{a})_{im}
\hat{c}^{\dagger}_i \hat{c}_m,
\end{equation}
and $\hat{A}_{coll}$ given by the Hermitean conjugate of 
$\hat{A}^{\dagger}_{coll}$. Here, $( O )_{ab}$ 
means the matrix elements of a single-particle operator $O$
between the nucleon states $\psi_a$ and $\psi_b$.
The symbols $\vec{a}^{\dagger}$ and $\vec{a}$ denote
the 
harmonic oscillator
``raising'' and ``lowering'' operators 
$\vec{a}^{\dagger} = \frac{1}{\sqrt{2}} \left[ (m\omega)^{1/2}
\vec{r} - i  (m\omega)^{-1/2} \vec{p} \right],
\quad
\vec{a} = \frac{1}{\sqrt{2}} \left[ (m\omega)^{1/2}
\vec{r} + i  (m\omega)^{-1/2} \vec{p} \right]$. 
It is seen that the role of the operators 
\begin{equation}
\label{BAREphonon}
\hat{A}^{\dagger}_{0} \equiv 
\sum_{mi} (\vec{\sigma}\vec{a}^{\dagger})_{mi}
\hat{c}^{\dagger}_m \hat{c}_i , \qquad
\hat{A}_{0} \equiv \sum_{mi} (\vec{\sigma} \vec{a})_{im}, 
\hat{c}^{\dagger}_i \hat{c}_m,
\end{equation}
is to create and destroy the correlated particle-hole excitations
with the total quantum numbers $J^{\pi} = 0^{-}$. Indeed,
the operator $a^{\dagger}$ raises the oscillator
principal quantum number $N$ by unity, while the role 
of the spin operator
in the internal product 
is to rotate properly 
the nucleon spins. The combined effect is
to remove particles from the orbital with the 
quantum numbers $[N,l,j]$ and place them in the orbital 
$[N+1,l \pm 1, j]$.
Thus, due to this construction the operator 
$\hat{A}^{\dagger}_{0}$ can only {\it create} the particle-hole pairs 
of $0^{-}$ type
($m$ and $i$ denote the states above and below the Fermi level, 
correspondingly). 
The effect of the {\it annihilation} operator 
$\hat{A}_{0}$ is just the opposite to that of $\hat{A}^{\dagger}_0$. 
 
The expressions for the phonon creation and destruction operators
$\hat{A}^{\dagger}$ and $\hat{A}$ are 
an ansatz for the
more general expression of collective RPA phonon operators 
given in Eq.(\ref{RPAphonon}).
The form of the
ansatz is based on the simple commutator properties for 
the operators $(\vec{\sigma} \vec{r})$ and $(\vec{\sigma} \vec{p})$
\cite{FV1}.
with the interactions
$V_0$ and $V_1$. (Note that if we substitute the interactions 
$V_0$ and $V_1$ by factorized expressions
$V_0 \rightarrow const (\vec{\sigma}\vec{r})_{ab}
(\vec{\sigma}\vec{r})_{cd}$,
and $V_0 \rightarrow const (\vec{\sigma}\vec{p})_{ab}
(\vec{\sigma}\vec{p})_{cd}$,
the ansatz (\ref{COLLECTIVEphonon}) 
will be an exact solution of the RPA equations (\ref{RPAequation})).

The use of Eq.(\ref{COLLECTIVEphonon}) allows us to 
get a closed system of RPA equations for the decoupled
collective phonon modes ($T=0,1$), and instead of 
system of a large number of coupled equations (\ref{RPAequation}) 
we obtain the simplified, approximate sytem of the
equations for the amplitudes $x$ and $y$ 
\begin{equation}
\label{RPA_COLL}
- E_{coll} \hat{A}^{\dagger}_{coll} \simeq 
 [ \hat{A}^{\dagger}_{coll}, H_0 ] + 
\langle [ \hat{A}^{\dagger}_{coll}, V_0 + V_1 ] \rangle,
\qquad \quad
- E_{sp} 
\hat{A}^{\dagger}_{sp} \simeq 
 [ \hat{A}^{\dagger}_{sp}, H_0 ]
\end{equation} 
From the first one of Eqs.(\ref{RPA_COLL}), the closed equations 
for the
collective phonon amplitudes $x$ and $y$ can be easily obtained 
using the following relations
\begin{equation}
\label{COMMUTATORS}
\langle [\hat{A}^{\dagger}_0, 
\hat{{\cal V}}_0
] \rangle = 
-\frac{\rho_0}{|U_0|}\frac{\omega}{2}(\hat{A}_0+\hat{A}^{\dagger}_0),
\qquad
\langle
[\hat{A}^{\dagger}_0, 
\hat{{\cal V}}_1
] \rangle = -2m\rho \omega (\hat{A}^{\dagger}_0-\hat{A}_0).
\end{equation}
where $\hat{{\cal V}}_0$ and $\hat{{\cal V}}_1$ denote the 
interaction terms 
$(\vec{\sigma}_1 \vec{\sigma}_2)
\delta(\vec{r}_1-\vec{r}_2)$
and 
$(\vec{\sigma}_1 \vec{\sigma}_2)
\{ \vec{p}_1, \{ \vec{p}_2, \delta(\vec{r}_1-\vec{r}_2) \} \}$
respectively expressed in the second quantization form
(the curly brackets
denote an anticommutator).
Hermitean conjugation gives the corresponding relations for the
commutators of $\hat{A}_0$ operators.
In the first one of Eqs.(\ref{COMMUTATORS}) 
we assumed the proportionality
of the nucleon density $\rho$ and the potential $U$:
$\rho \approx -\frac{\rho_0}{|U_0|} U$, where $\rho_0$ and $U_0$
are the values of the density and the potential in the center
of the nucleus.

All the remaining phonon degrees of freedom, marked by the subscript
'sp' in 
Eqs.(\ref{RPA_COLL}),  
are assumed here
to be essentially noncollective. The contribution from the interaction
term in the second one of Eqs.(\ref{RPA_COLL}) is dropped.
The frequencies of those  predominantly
single-particle modes are not shifted 
from the shell-model value $\omega$.

For the case of two kinds of nucleons ($p$ and $n$),
the collective phonons operators are sought in the form
${\hat A}^{\dagger}_{coll} = x_p {\hat A}^{\dagger}_0(p)+
x_n {\hat A}^{\dagger}_0(n)- y_p {\hat A}_0(p)-
y_n {\hat A}_0(n)$ instead of (\ref{COLLECTIVEphonon}).
From (\ref{RPA_COLL}),
we obtain
the following system of equations for the amplitudes 
$x_p,x_n,y_p$ and $y_n$
\begin{eqnarray}
\label{COLL_SYSTEM}
\left( \begin{array}{cccc}
E_T & 0 & -\omega (1+\frac{C\rho Z}{|U| A}g^{(+)}) 
& -\omega \frac{C\rho N}{|U| A}g^{(-)} \\ 
0 & E_T & -\omega \frac{C\rho Z}{|U| A}g^{(-)} 
& -\omega(1+ \frac{C\rho N}{|U| A}g^{(+)}) \\
\omega (1+\frac{C\rho m Z}{p_F^2 A}g_1^{(+)}) & 
\omega \frac{C\rho m N}{ p_F^2 A} g_1^{(-)} & -E_T & 0  \\ 
\omega \frac{C\rho m Z}{p_F^2 A}g_1^{(-)} & 
\omega \frac{C\rho m N}{ p_F^2 A} g_1^{(+)} & 0 & - E_T  
\end{array} \right)
\left( \begin{array}{c}
x_p-y_p  \\ 
x_n-y_n  \\
x_p+y_p  \\
x_n+y_n 
\end{array} \right)
\quad = \quad 0,
\end{eqnarray}
where $g^{(\pm)} = g \pm g'$ and $g^{(\pm)}_1 = g_1 \pm g'_1$.

It is convenient to analyze in detail the case of 
equal numbers of protons
and neutrons ($N=Z$).
From 
the consistency condition for (\ref{COLL_SYSTEM}) 
we find the two eigenfrequencies 
of the collective modes 
$E_{T=0}$ and $E_{T=1}$ 
to be
\begin{equation}
\label{FREQUENCY}
E_{T=0} = \omega \biggl[ \biggl(1+ \frac{C \rho m}{p_F^2} g_1 \biggr) 
\biggl(1+ \frac{C
\rho}{|U|} g \biggr) \biggr]^{1/2}, \quad
E_{T=1} = \omega \biggl[ \biggl(1+ \frac{C \rho m}{p_F^2} g'_1 \biggr)
 \biggl(1+ \frac{C
\rho}{|U|} g' \biggr) \biggr]^{1/2},
\end{equation} 

The eigenvectors of the system (\ref{COLL_SYSTEM})
are easily found using Eq.(\ref{FREQUENCY}),
in a form of phonon creation operators
\begin{eqnarray}
\label{PHONON_CREATORS}
\hat{A}^{\dagger}_{T=0} =
\frac{1}{2} 
\biggl[ \frac{\omega}{E_{T=0}(1+\frac{C \rho m}{p_F^2} g_1 ) 
{\cal N}_{coll}} \biggr]^{1/2}
\biggl\{ \biggl[ \frac{E_{T=0}}{\omega} + 1 + 
\frac{C \rho m}{p_F^2} g_1 \biggr]
(\hat{A}^{\dagger}_0(p) + \hat{A}^{\dagger}_0(n) )
\nonumber\\
-
\biggl[ - \frac{E_{T=0}}{\omega} + 1 + \frac{C \rho m}{p_F^2} g_1 
\biggr]
(\hat{A}_0(p) + \hat{A}_0(n) ) \biggr\},
\nonumber\\
\hat{A}^{\dagger}_{T=1} =
-\frac{1}{2} 
\biggl[ \frac{\omega}{E_{T=1}(1+\frac{C \rho m}{p_F^2} g'_1 ) 
{\cal N}_{coll}} \biggr]^{1/2}
\biggl\{
\biggl[\frac{E_{T=1}}{\omega} + 1 + \frac{C \rho m}{p_F^2} g'_1 \biggr]
(\hat{A}^{\dagger}_0(p) - \hat{A}^{\dagger}_0(n) )
\nonumber\\
+
\biggl[ - \frac{E_{T=1}}{\omega} + 1 + 
\frac{C \rho m}{p_F^2} g'_1 \biggr]
(\hat{A}_0(p) - \hat{A}_0(n) )
\biggr\},
\end{eqnarray}
where ${\cal N}_{coll}$ is
the colective phonon normalization
constant. The operators of the independent RPA modes should obey
the orthogonality and normalization conditions, thus
\begin{equation}
\label{COMMUTATORS_INT}
[ \hat{A}^{\dagger}_T, \hat{A}^{\dagger}_{T'} ]=0, \quad
\langle 0^+ [ \hat{A}_T, \hat{A}^{\dagger}_{T'} ] 0^+ \rangle = 
\delta_{T,T'}.
\end{equation}  
where the expectation value is taken over the uncorrelated 
ground state. 
The first set of Eqs.(\ref{COMMUTATORS_INT}) (orthogonality)
holds by construction, while the second one determines 
the normalization.
Substituting (\ref{PHONON_CREATORS}) into (\ref{COMMUTATORS_INT})
gives us the normalization relation
\begin{displaymath}
\langle 0^+ | [ \hat{A}_T, \hat{A}^{\dagger}_{T'} ] | 0^+ \rangle =
\delta_{T,T'} (R_p + R_n){\cal N}^{-1}_{coll}.
\end{displaymath}
Here, for a given kind of nucleons ($p$ or $n$), the quantity $R$ is 
\begin{displaymath}
R = \langle [ \hat{(\vec{\sigma} \vec{a})}, 
\hat{(\vec{\sigma} \vec{a}^{\dagger})}]
\rangle = \langle \sum_{i} (3 + 2 (\vec{\sigma} \vec{l})_{ii})
\hat{c}^{\dagger}_{i} \hat{c}_{i} \rangle
= 3 {\cal N} 
\end{displaymath}
where 
$\vec{l}$ is the single-particle 
orbital angular momentum.
The expectation value 
here 
is reduced to the summation over the occupied states $i$
and gives three times the
number of particles of  
given kind, ${\cal N}$. Note that for the spin saturated shell-model
ground state of a doubly magic nucleus, 
the term proportional to $(\vec{\sigma} \vec{l})_{ii}$
does not contribute. Therefore, the phonon normalization constant 
(for $N = Z = A/2$)
is
\begin{equation}
\label{NORM}
{\cal N}_{coll} = 3 A. 
\end{equation}

The matrix elements of the operators 
$(\vec{\sigma} \vec{r})_{p,n}$
and $(\vec{\sigma} \vec{p})_{p,n}$ between the 
correlated ground state $|0^+)$ and the excited single-phonon 
states $|0^-, T=0,1) = \hat{A}^{\dagger}_{T=0,1} |0^+)$
can be easily obtained by expressing the operators in terms
of the phonon creation and destruction operators 
$\hat{A}^{\dagger}_{T=0,1}$ and $\hat{A}_{T=0,1}$
[see Eqs.
(\ref{BAREphonon}),(\ref{PHONON_CREATORS})].
The results for the matrix elements of the 
isoscalar 
$W^{P}_0 = (0^+|(\vec{\sigma} \vec{p})  \hat{1} |0^-,T=0),
W^{PT}_0 = (0^+|(\vec{\sigma} \vec{r})  \hat{1} |0^-,T=0)$
and isovector parity violating operators
$W^{P}_1 = (0^+|(\vec{\sigma} \vec{p}) \hat{1} |0^-,T=1)$,
$W^{PT}_1 = (0^+|(\vec{\sigma} \vec{r}) \hat{1} |0^-,T=1)$
are found to be:
\begin{eqnarray}
\label{STRENGTHS}
W^{PT}_0 = 
\biggl( \frac{(1+\frac{C m \rho}{p^2_F}g_1) R}{2 m E_{T=0}} 
\biggr)^{1/2},
\qquad
W^{P}_0 = 
- i
\biggl( \frac{m E_{T=0} R}
{2 (1+\frac{C m \rho}{p^2_F}g_1) }
\biggr)^{1/2},
\nonumber\\
W^{PT}_1 = 
\biggl( \frac{(1+\frac{C m \rho}{p^2_F}{g'}_1) R}{2 m E_{T=1}} 
\biggr)^{1/2}
, \quad 
W^{P}_1 = 
- i 
\biggl( \frac{m E_{T=1} R}
{2 (1+\frac{C m \rho}{p^2_F}g'_1) } 
\biggr)^{1/2}.
\end{eqnarray}

To reveal the collective enhancement factors, we
compare these results to the 
matrix elements of the corresponding operators 
between shell-model states.
For the $(\vec{\sigma}\vec{p})$ and a given 1particle-1hole state,
$\hat{c}^{\dagger}_m \hat{c}_i |0^+ \rangle$,
we have the following relations 
\begin{equation}
\label{SP-SINGLE-PARTICLE}
\langle 0^+| \widehat{(\vec{\sigma} \vec{p})} \quad  
\hat{c}^{\dagger}_m \hat{c}_i | 0^+ \rangle 
= -i\left(\frac{m \omega}{2}\right)^{1/2} (\vec{\sigma}\vec{a})_{mi} 
\quad \sim 
\quad \left(\frac{m \omega}{2}\right)^{1/2} 
\sqrt{N_F} \sim \left(\frac{m \omega}{2}\right)^{1/2} (3 A)^{1/6}
\end{equation}
where the last expressions indicate the order of magnitude.
$N_F$ denotes the principal oscillator quantum number at
the Fermi energy of a doubly magic nucleus. We used 
in (\ref{SP-SINGLE-PARTICLE}) the
fact that $N_F \sim (3A)^{1/3}$. 
Similarly:
\begin{equation}
\label{SR-SINGLE-PARTICLE}
\langle 0^+| \widehat{(\vec{\sigma} \vec{r})} \quad  
\hat{c}^{\dagger}_m \hat{c}_i | 0^+ \rangle 
= \left(\frac{1}{2m \omega}\right)^{1/2} (\vec{\sigma}\vec{a})_{mi} 
\quad \sim 
\left(\frac{1}{2m \omega}\right)^{1/2}\sqrt{N_F} \sim 
\left(\frac{1}{2m \omega}\right)^{1/2} (3A)^{1/6}
\end{equation}

Comparing the matrix elements
(\ref{STRENGTHS}) to the single-particle
ones (\ref{SP-SINGLE-PARTICLE}), (\ref{SR-SINGLE-PARTICLE}) 
and taking into account Eq.((\ref{FREQUENCY}),
we see that the matrix elements of the P-odd potentials connecting the
correlated ground state to the single-phonon states
are enhanced by the factors 
\begin{equation}
\label{ENHANCEMENT}
F^{P}_{coll} \approx 
\frac{|(0^+| (\vec{\sigma} \vec{p}) |0^-,T=0)|}
{|\widehat{(\vec{\sigma} \vec{p})}_{mi}|} 
\sim \quad 
\sqrt{\frac{\omega}{E_0}(1+\frac{C \rho}{|U|}g)} \quad (3A)^{1/3}.
\end{equation}
and
\begin{equation}
\label{ENHANCEMENT2}
F^{PT}_{coll} \approx
\frac{|(0^+| (\vec{\sigma} \vec{r}) |0^-,T=0)|}
{|\widehat{(\vec{\sigma} \vec{r})}_{mi}|}  
\sim \quad 
\sqrt{\frac{\omega}{E_0}(1+\frac{C m \rho}{p_F^2}g_1)} 
\quad (3A)^{1/3}.
\end{equation}
Analogous expressions can be written for the $T=1$ matrix elements 
by substituting $g' (g'_1)$ instead of $g (g_1)$ in 
(\ref{ENHANCEMENT},\ref{ENHANCEMENT2}).

The total collective enhancement factor $F_{coll}$ has a natural 
form typical 
for the low-energy collective modes.
On the right hand side of Eqs.(\ref{ENHANCEMENT}),(\ref{ENHANCEMENT2}), 
the second factor is the collective ``phase volume'' enhancement,
$F_{pv}$,
which is roughly equal to the square root of
the number of single-particle constituents of the phonon, i.e.,
approximately the 
number of nucleons in the upper major shell, ${\cal A}_F$),
\begin{displaymath}
F_{pv} \sim \sqrt{{\cal A}_F} \sim N_F \sim (3A)^{1/3}. 
\end{displaymath}

The first factor
in Eqs.(\ref{ENHANCEMENT}),(\ref{ENHANCEMENT2}) 
is the ``adiabaticity parameter'' \cite{Z}. It reflects 
the change of the frequency of the collective motion compared to 
the frequency of single-particle motion $\omega$
\begin{displaymath}
F_{ad} \sim \sqrt{\frac{\omega}{E_0}(1+\frac{C \rho}{|U|}g)} 
\qquad (P - odd),
\qquad   
F_{ad} \sim 
\sqrt{\frac{\omega}{E_0}(1+\frac{C m \rho}{p_F^2}g_1)}
\qquad (P-,T - odd).
\end{displaymath}
Note that for the low-lying
$0^-$-mode ($T=0$), $F_{ad}>1$ gives additional enhancement.

We turn now to a quantitative 
discussion 
of the above results. 
Using the conventional
constants of the Landau-Migdal interaction (\ref{V0}),(\ref{V1}) 
$g=0.575, g'=0.725$,  $g_{1}=-0.5$ and 
$g'_{1}=-0.26$ 
(see e.g. Refs.\cite{Migdal},\cite{GB},\cite{FSTA},\cite{KS}),
we find from Eq.(\ref{FREQUENCY}):
\begin{equation}
\label{VALUES}
E_{T=0} = 1.03 \omega , \qquad E_{T=1} = 1.20 \omega .
\end{equation}
The dependence of $E_{T}$ on the total strength of the interaction
constant is shown on Fig.1.
It is seen that due to the negative value of  
$g_1$, $g'_1$,
the effect of the
velocity dependent interaction is to reduce the frequency
of the isoscalar mode pushing it towards instability \cite{FV2}.
However, the velocity independent component acts in the opposite
direction. 
The excitation energy of the isovector mode that is governed by the
primed constants tends to increase with the interaction strength.
(These tendencies are in agreement with those found in 
numerical studies \cite{AK} and with results 
for the meson exchange interactions \cite{FV1}). 
The combined and opposite action of $V_0$ and $V_1$ on $E_{T}$ tends to
moderate energy shift from $\omega$, so the values (\ref{VALUES})
do not mean that the correlations in the ground state are small.
The collective enhancement of the matrix elements (\ref{STRENGTHS})
are $F^P \sim 10-15$ for the above values of $g$.

We should stress that it is not the prime purpose of this work to
achieve agreement with experiment. 
The mean field that provides the basis of the single-particle
energies is not treated in this work. Therefore effects of
consistency between the $p-h$ interaction
and the mean field \cite{GB} are not taken into account.
Also, we do not include the one-body spin-orbit interaction 
which for $0^-$ mode could be important especially 
in lighter nuclei \cite{TCZ}.

As we mentioned, the velocity dependent Landau-Migdal interaction
can be viewed as a model of a more realistic momentum-dependent 
$\pi$- and $\rho$-meson exchange interactions.
Using $V_{\pi+\rho}$ of Eq.(\ref{PIRHO}) as a residual interaction
within the above formalism we find 
\begin{eqnarray}
\label{FREQUENCY_PI_RHO}
E_{T=0} = \omega \biggl[ \biggl(1 - \frac{3 k}{2} \biggr) 
\biggl(1+ \frac{C
\rho}{|U|} g - 3 \xi \biggr) \biggr]^{1/2}, 
\quad
\nonumber\\
E_{T=1} = \omega \biggl[ \biggl(1 + \frac{k}{2} \biggr) 
\biggl(1+ \frac{C
\rho}{|U|} g' - 2 \phi + \xi \biggr) \biggr]^{1/2},
%
\end{eqnarray}
Here, the constants $k$ and $\phi$ and $\xi$ are related to
the meson-nucleon coupling strenghts: 
$k= 2 \rho_0 m q $ results from the evaluation of the exchange terms
\cite{FV1} with $q=6 \pi (\frac{f_{\pi}^{2}}
{m_{\pi}^{4}}
W_{\pi}-\frac{4}{3}\frac{f_{\rho}^{2}}
{m_{\rho}^{4}}
W_{\rho})
$ 
and  the nonlocality factors $W$ 
($W_{\pi,\rho} \rightarrow 1$ for 
$m_{\pi,\rho} \rightarrow \infty$)
are 
$W_{\pi}=0.11$,
$W_{\rho}=0.69$ 
for the pion and $\rho$-meson
correspondingly \cite{FV1}. 
$\phi$ results from
the direct terms, 
$\phi = \frac{v}{m \omega^2}$, 
$v= - \frac{4 \pi}{3} \frac{\rho_0}{|U_0|} m \omega^2
(\frac{f^2_{\pi}}{m^2_{\pi}}+ 2\frac{f^2_{\rho}}{m^2_{\rho}})
$,
while the term
$\xi$ is due to 
the exchange terms
and $\xi=\frac{2 \pi}{3}\frac{\rho_0}{|U_0|} (-\frac{f_{\pi}^{2}}
{m_{\pi}^{2}}
{W'}_{\pi}
+2\frac{f_{\rho}^{2}}
{m_{\rho}^{2}}
{W'}_{\rho})$ 
where 
${W'}_{\pi}= 0.56$ and ${W'}_{\rho}= 0.71$.
The pion constant is equal $f^2_{\pi}=0.08$, while for the $\rho$-meson
the coupling constant is in range from $f^2_{\rho}=1.86$ 
(``weak $\rho$-meson'')
to $f^2_{\rho}=4.86$ (``strong $\rho$-meson'') \cite{JKWB},\cite{BB}.

The phonon operators $\hat A^{\dagger}_{T}$, $\hat A_{T}$ are 
given now by the same expressions 
as before [Eq.(\ref{PHONON_CREATORS})] but with the replacements
$1+\frac{C \rho m}{p_F^2} g_1 \rightarrow 
1 - 3k/2  $ and
$1+\frac{C \rho m}{p_F^2} g'_1 \rightarrow 
1 + k/2 $ and 
with 
$E_{T=0,1}$ given by (\ref{FREQUENCY_PI_RHO}).
Correspondingly, the matrix elements of the symmetry violating
operators 
\begin{eqnarray}
\label{STRENGTHS_PIRHO}
W^{PT}_0 = 
\biggl( \frac{(1
- 3  k/2 ) R}{2 m E_{T=0}}
\biggr)^{1/2}
, \quad 
W^{P}_0 = 
- i
\biggl( \frac{m E_{T=0} R}
{2 (1
- 3k/2 ) }
\biggr)^{1/2},
\nonumber\\
W^{PT}_1 = 
\biggl( \frac{(1
+ k/2 ) R}{2 m E_{T=1}} 
\biggr)^{1/2}
, \quad 
W^{P}_1 = 
- i 
\biggl( \frac{m E_{T=1} R}
{2 (1
+ k/2 ) } 
\biggr)^{1/2}
.
\end{eqnarray}

The results for the frequencies are plotted in 
Fig.1. 
It is seen that the effect of the meson exchange interaction is to push 
down the isoscalar frequency $E_{T=0}$, mainly due to the $\pi$-meson
contribution. 
At the same time it shift upwards the isovector  
frequency. Similar behaviour of the $E_{T=1}$ was obtained in
a more realistic RPA calculations in Ref.\cite{AK}.
Fig.2a shows $E_T$ as a function of $f^2_{\rho}$.
Instability of the $0^-,T=0$ occurs for the ``weak $\rho$-meson'',
as was discussed in \cite{FV1}
using general properties of the linear response \cite{PN}.
It is interesting that the collapse occurs
also for the ``strong'' $\rho$-meson, while 
the stable energy range $E_{0} \simeq (0.2-0.4)\omega$ occurs for 
$1.86 < f^2_{\rho} <4.86$.  
Fig. 2b presents the enhancement factors $F^P$ and $F^{PT}$
of the weak matrix elements 
modified according to (\ref{STRENGTHS_PIRHO}).
The collective enhancement for the P-odd matrix elements
rises from $10-20$ to considerably larger values as $E^2_{T=0} 
\rightarrow
0$.

The enhancement of the P-odd matrix elements connecting the
ground state to the single-phonon states
found here
can result in 
observable P-effects [e.g.,  in the electromagnetic transitions (multipole
mixing) \cite{AH}].
For the P-odd ($T=0$) matrix element, where the enhancement is maximum,
the quantity $W^P_{coll}$ can reach, for example for $^{206}Pb$, a  value 
of
\begin{displaymath}
W^P_{coll} = \quad (0^+| g^W_p \frac{G}{2 m \sqrt{2}}
\{ (\vec{\sigma}_p \vec{p}_p), \rho \} + 
g^W_n \frac{G}{2 m \sqrt{2}}
\{ (\vec{\sigma}_n \vec{p}_n), \rho \} | 0^-,T=0)
\quad \simeq \quad 20 - 30 eV,
\end{displaymath}
for the realistic weak interaction strengths $g_p \simeq 4$ and $g_n \simeq 1$
(see, e.g., \cite{FTS}) consistent with the standard parameters 
of the two-body weak interaction \cite{DDH},\cite{AH} 
($G$ is the Fermi constant).

The collective of $0^-$ states has been already 
considered in the doorway mechanism approach \cite{A},\cite{AB} 
in the problem of P-odd mixing in compound nuclei.
The enhancement in the parity violating matrix element due
to the collectivity of the $0^-$ was one of the assumptions in the above
works. The results of the present work confirm their idea and the 
collectivity of the P-odd matrix element is a natural result 
of the RPA treatment of the $0^-$ states.

One of us (N. A.) is grateful to B. Mottelson and 
L. Zamick for discussions.
O.V. is grateful to V. G. Zelevinsky for discussions of various 
problems of collective states. The authors acknowledge  
the participation at the ECT Trento workshop.
The work was supported by the US-Israel Binational 
Science Foundation and
the Israeli Science Foundation.

\noindent

\newpage

\begin{center}
{\large {\bf Figure captions}}
\end{center}

{\bf Fig. 1.} 
The $0^-$ energies $E_{T=0}$ and $E_{T=1}$ 
[Eq.(\ref{FREQUENCY}),
in units of $\omega$] as a function of the
strength of the Landau-Migdal interaction $C$
[taken in units of the standard value (Eq.\ref{V0},\ref{V1})]
(solid curves). The dashed curves show $E_{T=0}$ and $E_{T=1}$
as functions of the pion coupling constant $f^{2}_{\pi}$ 
(the scale is given in the upper line of the box) for the $\rho$-meson
constant $f^{2}_{\rho}$ taken in the ``weak meson'' limit.
The collapse of the RPA frequency $E_{T=0}$ occurs when the actual 
value of the pion constant $f^{2}_{\pi} \approx 0.08$ is used.

\vspace{1cm}
{\bf Fig. 2.   a)}
Energies of the $0^-$ states $E_{T=0}$ and $E_{T=1}$ 
[Eq.(\ref{FREQUENCY_PI_RHO})],
in units of $\omega$ are plotted as a function of the
$\rho$-meson strength $f^2_{\rho}$
calculated for the standard value of the $\pi$-meson constant.
Collapse points in $E_{T=0}$ occur at values of $f^2_{\rho}$ 
approximately equal
to the ``strong'' and ``weak'' $\rho$-meson couplings.
 
{\bf b)}
The enhancement factors $F^P$ (crosses) and $F^{PT}$ (solid lines)
[Eq.(\ref{ENHANCEMENT})] as function of $f^2_{\rho}$ (for the standard
value
of $f^2_\pi$).
\end{document}